\documentclass[manuscript]{acmart}

\AtBeginDocument{%
  \providecommand\BibTeX{{%
    \normalfont B\kern-0.5em{\scshape i\kern-0.25em b}\kern-0.8em\TeX}}}

\setcopyright{acmcopyright}
\copyrightyear{2021}
\acmYear{2021}

\acmConference[RecSys '21]{RecSys '21: The ACM Conference Series on Recommender Systems}{27th September - 1st October 2021}{Amsterdam, Netherlands}

\begin{document}

\title[A new sequential dataset for recommender systems]{A new sequential dataset logging interactions, all viewed items and click responses/no-click for recommender systems research}

\author{Simen Eide}
\email{simen.eide@finn.no}
\orcid{1234-5678-9012}
\affiliation{%
  \institution{University of Oslo and FINN.no}
  \city{Oslo}
  \country{Norway}}
  
\author{Arnoldo Frigessi}
\email{arnoldo.frigessi@medisin.uio.no}
\affiliation{%
  \institution{University of Oslo}
  \city{Oslo}
  \country{Norway}}
  
\author{Helge Jenssen}
\email{helge.jenssen@finn.no}
\affiliation{%
  \institution{FINN.no}
  \city{Oslo}
  \country{Norway}}

\author{David S. Leslie}
\email{d.leslie@lancaster.ac.uk}
\affiliation{%
  \institution{University of Lancaster}
  \city{Lancaster}
  \country{United Kingdom}}

\author{Joakim Rishaug}
\email{joakim.rishaug@finn.no}
\affiliation{%
  \institution{FINN.no}
  \city{Oslo}
  \country{Norway}}
  
\author{Sofie Verrewaere}
\email{Sofie.verrewaere@schibsted.com}
\affiliation{%
  \institution{Schibsted ASA}
  \city{Oslo}
  \country{Norway}}

\renewcommand{\shortauthors}{Eide, et al.}


%

\maketitle

\section{Introduction}
\begin{figure}[h]
  \centering
  \includegraphics[width=\linewidth]{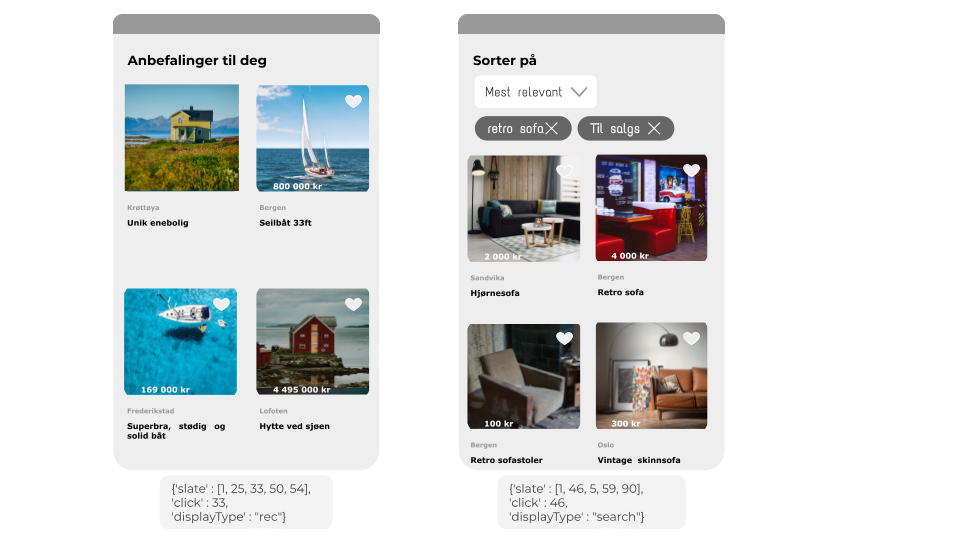}
  \caption{An example screenshot of the logged slates as they appear on FINN.no. Left shows a recommendation slate. Right shows the resulting slate of a search query. In both cases four items are exposed to the user. We also provide examples (below) of how these slates would have been logged in the dataset: "slate" is a list of the viewed items including the special item of "no click" (which is the index 1), "click" is the index of the clicked item, and "displayType" encodes whether it is a search or recommendation slate.}
  \Description{An example screenshot of the recommendation system at FINN.no}
  \label{fig:frontpage}
\end{figure}

We present a novel recommender systems dataset that records the sequential interactions between users and an online marketplace. The users are sequentially presented with both recommendations and search results in the form of ranked lists of items, called slates, from the marketplace (figure \ref{fig:frontpage}). 
The dataset includes the presented slates at each round, whether the user clicked on any of these items and which item the user clicked on.
Although the usage of exposure data in recommender systems is growing \citep{Ie2019a, Chen2019, Eide2021DynamicSampling, Edwards2018}, to our knowledge there is no open large-scale recommender systems dataset that includes the slates of items presented to the users at each interaction.
As a result, most articles on recommender systems do not utilize this exposure information.
Instead, the proposed models only depend on the user's click responses, and assume that the user is exposed to all the items in the item universe at each step, often called uniform candidate sampling (\citep{Hu2008, Hidasi2016, Covington2016}).
This is an incomplete assumption, as it takes into account items the user might not have been exposed to. This way items might be incorrectly considered as not of interest to the user.
Taking into account the actually shown slates allows the models to use a more natural likelihood, based on the click probability given the exposure set of items, as is prevalent in the bandit and reinforcement learning literature \citep{Lattimore2019, Barto2015, Zhao2018, Ie2019, McInerney2018}.
\cite{Eide2021DynamicSampling} shows that likelihoods based on uniform candidate sampling (and similar assumptions) are implicitly assuming that the platform only shows the most relevant items to the user.
This causes the recommender system to implicitly reinforce feedback loops and to be biased towards previously exposed items to the user.

We hope that the release of this dataset can assist researchers to construct more realistic user preference models by utilizing the exposed slates and contribute to more robust offline evaluation criteria using for example importance sampling \citep{joachims}.
The dataset along with quickstarts and tutorials are available at \href{https://github.com/finn-no/recsys-slates-dataset}{https://github.com/finn-no/recsys-slates-dataset}.

\subsection{Related datasets}
There are some datasets on ad display or information retrieval \citep{criteo, Rekabsaz2021}. However, these datasets have some shortcomings. Failing to identify the user or lacking a sufficient number of interactions per user cause them to be unfitting for benchmark analyses. 
There are also simulation environments that allow interaction with synthetic users \citep{Iey2019RECSIM:Systems}, showing the need for real-world datasets.

\subsection{About the marketplace}
FINN.no is the leading marketplace in Norway for real estate, job offerings, vehicles and general merchandise. It is also one of the largest websites in the country.
Approximately 20\% of the clicks on item listings come from its recommender systems, and the system recommends around 90 million items each day. FINN.no is part of the media group Schibsted ASA.

\subsection{Detailed description of the dataset}
The present dataset was collected over 30 days on the marketplace FINN.no.
The dataset contains 37.4 million logged search or recommendation user interactions.
For each item, we log all the items presented in the slate and the user response.
In addition, we provide a text string attribute for each item containing its categories.
Approximately 70\% of the interactions originate from search queries, the remaining 30\% from recommendations.
24.4\% of the interactions resulted in the user not clicking on any item.
The dataset consists of 2.3 million users and 1.3 million unique items.
We provide more dataset statistics in Table \ref{tab:dataset-stats}.

\paragraph{Interaction data}
The dataset contains the interaction types, the presented slates and the user responses.
The interaction type identifies whether it was a search query or the recommendation system that presented the slate.
The presented slate contains all items that were displayed to the user during each interaction (the full scroll length by the user), but not necessarily in the observed order.
For example, in figure \ref{fig:frontpage}, we would have logged 4 items in both cases.
Finally, the dataset also includes the user response which may be one of the items in the presented slate or the action of not clicking on any of the items (\textit{no-click}).

\begin{table}
  \caption{Dataset statistics}
  \label{tab:dataset-stats}
  \begin{tabular}{ccl}
    \toprule
    Description & Value \\
    \midrule
    Number of interactions  & 37.5m \\
    Length of time period of data collection & 30 days \\
    Unique Items            & 1.3m\\
    Unique item groups      & 290 \\
    Unique Users            & 2.3m\\
    Average number of interactions per user & 16.4 \\
    Rate of interactions with \textit{no-click} & 24.4\% \\
    Average Slate length    & 10.1 items \\
    Percentage of search produced slates & 69.7\% \\
   Percentage of recommendation produced slates & 30.3\% \\
  \bottomrule
\end{tabular}
\end{table}

\paragraph{Item attributes}
The items belong to 290 unique groups that are constructed using a combination of categorical information and the geographical location describing  where the item is made available in the marketplace. 
Each group label describes this information using a text string.
For example, the group \textit{BAP, antiques, Trøndelag} is interpreted as belonging to the general merchandise ("Bits And Pieces") category, in the antique category and being sold in the county of \textit{Trøndelag}.
Similarly, the group \textit{MOTOR, Sogn og Fjordane} is interpreted as the item belonging to the motor category, and sold in the county \textit{Sogn og Fjordane}. 
Note that in this group the categorical information is omitted, to keep the groups sufficiently large for anonymization requirements.

\paragraph{Processing and anonymization}
The published dataset has been processed to anonymize its content and to enhance computation.
Due to the abundant amount of \textit{no-click} signals (which were 76.3\% of the total number of interactions), the \textit{no-click} interactions were first uniformly down-sampled to 10\% of the original number of interactions.
Users with less than 10 interactions during the data collection period have been removed.
If a user had more than 20 interactions, only the 20 first occurring interactions are logged.
The maximum number of items considered in a slate is capped at 25 items. This capping impacts only 5\% of the slates.
If the user clicked on any of the items that were removed, we removed the whole interaction from the dataset.
These simplifications only affect a small portion of the dataset and do not, in our view, alter the main contribution of the dataset.

\paragraph{Train/Validation/Test Split}
We propose a way to split the dataset into training, validation and test dataset.
This is also the split used to evaluate the models in \cite{Eide2021DynamicSampling}.
Let all interactions for 90\% of the users be placed in the training dataset. For the remaining 10\% of the users, we place 5\% in each validation and test dataset. 
However, the first five interactions per user are placed in the training dataset.
This allows a recommender system to estimate some user-specific parameters based on limited knowledge of the user.
These splits are implemented as \href{https://github.com/finn-no/recsys-slates-dataset/blob/main/code/dataset.py}{\textit{Pytorch DataLoaders} in the repository} for easy application.

\paragraph{Limitations}
We have explicitly not included some data fields that may have been relevant for recommender systems research, mostly due to privacy concerns.
For example, including a full description of items, the text queries that were used in search slates and exact timestamps of all interactions have been excluded because we did not find a way to share this information without the risk of exposing private information.
We also limited the total number of interactions for the same reason.

\begin{acks}
We acknowledge funding from BigInsight (Norwegian Research Council project number 237718) 
and the joint Industrial PhD project between FINN.no, University of Oslo and the Norwegian Research Council (Norwegian Research Council project number 294330).
This research is also part of the strategic partnership between BigInsight at the University of Oslo and the STORi Centre for Doctoral Training at Lancaster University.
\end{acks}

\bibliographystyle{ACM-Reference-Format}
\bibliography{references}

\end{document}